\documentclass[letterpaper]{article}

\usepackage{natbib,alifeconf}  
\usepackage{gensymb}
\usepackage{amsmath}
\usepackage{array}
\usepackage{subcaption}
\usepackage{multirow}
\usepackage{rotating}
\usepackage{fancyhdr}

\newcommand{\eqnref}[1]  {equation~(\ref{#1})}
\newcommand{\figref}[1]  {Fig.~\ref{#1}}
\newcommand{\tabref}[1]  {table~\ref{#1}}

\renewcommand{\thispagestyle}[1]{} 

\title{An active inference implementation of phototaxis \thanks{This work is licensed to the public under a Creative Commons Attribution- NonCommercial-NoDerivatives 4.0 license (international): http://creativecommons.org/licenses/by-nc-nd/4.0/}}
\author{Manuel Baltieri, Christopher L. Buckley \\
\mbox{}\\
Evolutionary and Adaptive Systems Group, Department of Informatics, \\ University of Sussex, Brighton, UK \\
m.baltieri@sussex.ac.uk} 

\begin{document}
\maketitle

\begin{abstract}
Active inference is emerging as a possible unifying theory of perception and action in cognitive and computational neuroscience. On this theory, perception is a process of inferring the causes of sensory data by minimising the error between actual sensations and those predicted by an inner \emph{generative} (probabilistic) model. Action on the other hand is drawn as a process that modifies the world such that the consequent sensory input meets expectations encoded in the same internal model. These two processes, inferring properties of the world and inferring actions needed to meet expectations, close the sensory/motor loop and suggest a deep symmetry between action and perception. In this work we present a simple agent-based model inspired by this new theory that offers insights on some of its central ideas. Previous implementations of active inference have typically examined a ``perception-oriented" view of this theory, assuming that agents are endowed with a detailed generative model of their surrounding environment. In contrast, we present an ``action-oriented" solution showing how adaptive behaviour can emerge even when agents operate with a simple model which bears little resemblance to their environment. We examine how various parameters of this formulation allow phototaxis and present an example of a different, ``pathological" behaviour.
\end{abstract}

\section{Introduction}
Brains must operate in an uncertain world, with noisy sensors that provide only incomplete and often ambiguous information. Recent developments in cognitive and computational neuroscience have suggested that the brain meets this challenge by operating as a Bayesian inference machine. This idea is usually traced back to work by Helmholtz \citep{von1867handbuch} and his theory of unconscious inference. On this view, perception is cast as an ongoing process of updating an inner \emph{generative} model so that it can best recapitulate (or ``predict") noisy and ambiguous incoming sensory input and thus infer the hidden (i.e. not directly accessible by the brain) causes of such data \citep{dayan1995helmholtz, rao1999predictive, knill2004bayesian, friston2006free, clark2013whatever, hohwy2013predictive, bogacz2015tutorial, buckley2017free}.

Predictive Coding models represent one concrete instantiation of this inferential process and have been used, for instance, to account for the neural dynamics underlying perception in the visual cortex \citep{rao1999predictive}. These ideas have been significantly extended by the Free Energy Principle (FEP) \citep{friston2006free, Friston2010nature}, which also provides a mechanistic account of action within the same framework. Specifically, under the FEP, while perception (``perceptual inference") is a process of updating an inner model to best account for sensory data, actions change the world to make sensory input better accord with predictions made by the same model (``active inference"). It has been suggested that the interplay between these two processes acting to satisfy a generative model that encodes constraints (``priors") conducive to an agent's survival can form the foundations of adaptive behaviour \citep{friston2012dark}.

The vast majority of models implemented using this framework have assumed that agents are endowed with a detailed generative model of their surrounding environment. These ``perception-oriented" approaches subordinate motor actions to the accurate and comprehensive perception of the environmental causes of sensory data \citep{hohwy2013predictive} and thus have often brought the FEP and active inference into direct conflict with more enactivist views of cognition \citep{clark2015radical, bruineberg2016anticipating, allen2016cognitivism}. In contrast, others have suggested that complex adaptive behaviour could  emerge from the interplay between an agent acting on the basis of simpler, more frugal generative models and the environment \citep{clark2015radical}. This ``action-oriented" perspective could underpin a more ecological and embodied reading of the FEP \citep{seth2014cybernetic, clark2015radical, bruineberg2016anticipating, allen2016cognitivism}.

In this work we show an example of a simple wheeled agent performing phototaxis under active inference and present it  as a proof of principle of an ``action-oriented" reading on the FEP. We also examine how phototaxis depends on various parameters settings and how this could be used as a generic model of different emergent behaviours.
\looseness=-1

\section{The Free Energy Principle (FEP) and Active Inference}
Bayesian accounts of perception hold that a central goal of agentive systems is to infer the hidden environmental causes of sensory data \citep{knill2004bayesian}. Formally, this can be written as a process of Bayesian inference in terms of the causes $x$ of sensory input $\rho$:
\begin{eqnarray}
P(x | \rho) = \frac{P(\rho | x) P(x)}{P(\rho)}
\end{eqnarray}
where $P(x | \rho)$ is the \emph{posterior} probability of hidden causes $x$ given observed sensory data $\rho$. $P(\rho | x)$ is the \emph{likelihood}, corresponding to the organism's assumptions about how sensory input $\rho$ relates to hidden causes $x$. $P(x)$ is the \emph{prior}, encoding the agent's ``beliefs" about hidden causes before it receives $\rho$ and $P(\rho)$ is the \emph{marginal likelihood}, a normalisation factor obtained by marginalising $P(\rho | x)$ over all possible causes $x$. To calculate the posterior probability it is necessary to evaluate the marginal likelihood  (also called ``surprisal", \cite{Friston2010nature}) $P(\rho)$, which is often difficult if not practically intractable \citep{bishop2006pattern, buckley2017free}. Variational Free Energy represents an approximate technique for Bayesian inference \citep{bishop2006pattern} that has been argued to be compatible with a neurally plausible implementation of this Bayesian scheme \citep{Friston2008c}. The method involves optimising an auxiliary probability density $Q(x)$, referred to as a \emph{recognition density}, so that it becomes a good approximation of the posterior $P(x | \rho)$. This can be achieved by minimising a measure of the difference between these two densities, quantified as the Kullback-Leibler (KL) divergence \citep{kullback1951information}
\begin{eqnarray}
D_{KL}(Q(x)||P(x | \rho)) = \int Q(x) \ln \frac{Q(x)}{P(x | \rho)} dx 
\label{eq:KL}
\end{eqnarray}
and while we cannot evaluate this expression directly since it still involves the unknown posterior, we can rewrite it as
\begin{eqnarray}
D_{KL}(Q(x)||P(x | \rho)) = F + \ln P(\rho) \label{eq:KLpost}
\end{eqnarray}
where we defined the ``variational free energy" as 
\begin{eqnarray}
F  \equiv \int Q(x) \ln \frac{Q(x)}{P(x,\rho)} dx
\label{eq:freeEnergyKL}
\end{eqnarray}
Unlike \eqnref{eq:KL}, the free energy $F$ can be evaluated because it only involves the recognition density, which we are free to specify, and a model of the world dynamics in terms of a prior and a likelihood which we assume an agent has, i.e. $P(x,\rho) = P(\rho | x) P(x)$. The second term on the right-hand side in \eqnref{eq:KLpost} is independent of the recognition density $Q(x)$ (it only depends on sensory input $\rho$). Thus, minimising \eqnref{eq:freeEnergyKL} with respect to $Q(x)$ will minimise the KL divergence between the recognition density and the true posterior. The result of this minimisation will make $Q(x)$ approach the true posterior $P(x | \rho)$. Optimising free energy for arbitrary recognition densities can be complex, so a common assumption is to restrict the form of $Q(x)$ to a tightly peaked Gaussian distribution, i.e. the Laplace approximation \citep{friston2006free, Friston2008a, bogacz2015tutorial, buckley2017free}. Variables $x$ are then replaced by parameters $\mu_x$ representing the first order sufficient statistics (i.e. the mean(s)) of this Gaussian distribution. Effectively, $\mu_x$ represent a parametrisation of an agent's beliefs or best guesses of the most likely causes $x$. It can be shown \citep{Friston2008a, bogacz2015tutorial, buckley2017free} that under these assumptions the free energy term simplifies to
\begin{eqnarray}
F = - \ln P(\rho, \mu_x) + constants
\label{eq:freeEnergyLaplace}
\end{eqnarray}
where $P(\rho, \mu_x) = P(\rho| \mu_x) P(\mu_x)$ is the generative density comprising of a likelihood $P(\rho| \mu_x)$ and a prior $P(\mu_x)$ in terms of parametrised beliefs about hidden causes $\mu_x$. In sum, changing beliefs $\mu_x$ to minimise the free energy $F$, constrained by sensory data $\rho$, makes $\mu_x$ the best guess/estimate of hidden causes $x$.

Under this framework it is suggested that perception is implemented as the minimisation of free energy with respect to beliefs $\mu_x$ following a gradient descent scheme:
\begin{eqnarray}
\dot{\mu}_x = - \frac{\partial F}{\partial \mu_x}
\label{eq:perceptionMin}
\end{eqnarray}
This equation updates $\mu_x$ and converges when the minimum of the free energy $F$ is reached, i.e. when $\frac{\partial F}{\partial \mu_x} =0$.

In contrasts to perception, action is defined as a process of changing the world such that sensory data better accords with predictions of the generative model (\figref{fig:diagram}). Specifically, in terms of the formalism presented above, while perception minimises the first term of \eqnref{eq:KLpost}, action optimises the second one by updating sensations $\rho$. To achieve this, an agent must know (or at least have an approximation of) how $\rho$ depend on motor action $a$ (i.e. $\rho = f(a)$) \citep{Friston2010biocyb, buckley2017free}. Given this, action can similarly be cast as a gradient descent on the free energy with respect to the variable $a$
\begin{eqnarray}
\dot{a} = - \frac{\partial F}{\partial a} = - \frac{\partial F}{\partial \rho} \frac{\partial \rho}{\partial a}
\label{eq:actionMin}
\end{eqnarray}
Thus action and perception can be described as the minimisation of the same quantity, with the simultaneous implementation of both processes closing the action-perception loop. \looseness=-1

\section{The Model}
\begin{figure}[ht!]
\begin{subfigure}{.5\textwidth}
  \centering
  \includegraphics[width=1.7in]{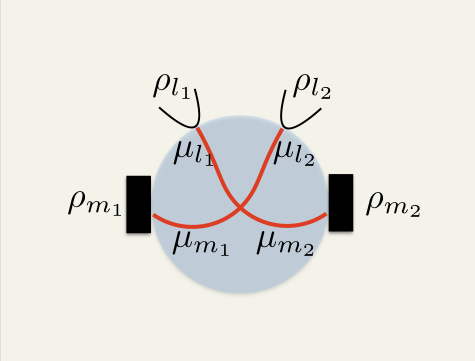}
\caption{}
\label{fig:vehicle}
\end{subfigure}
\begin{subfigure}{.5\textwidth}
  \centering
  \includegraphics[width=.9\textwidth]{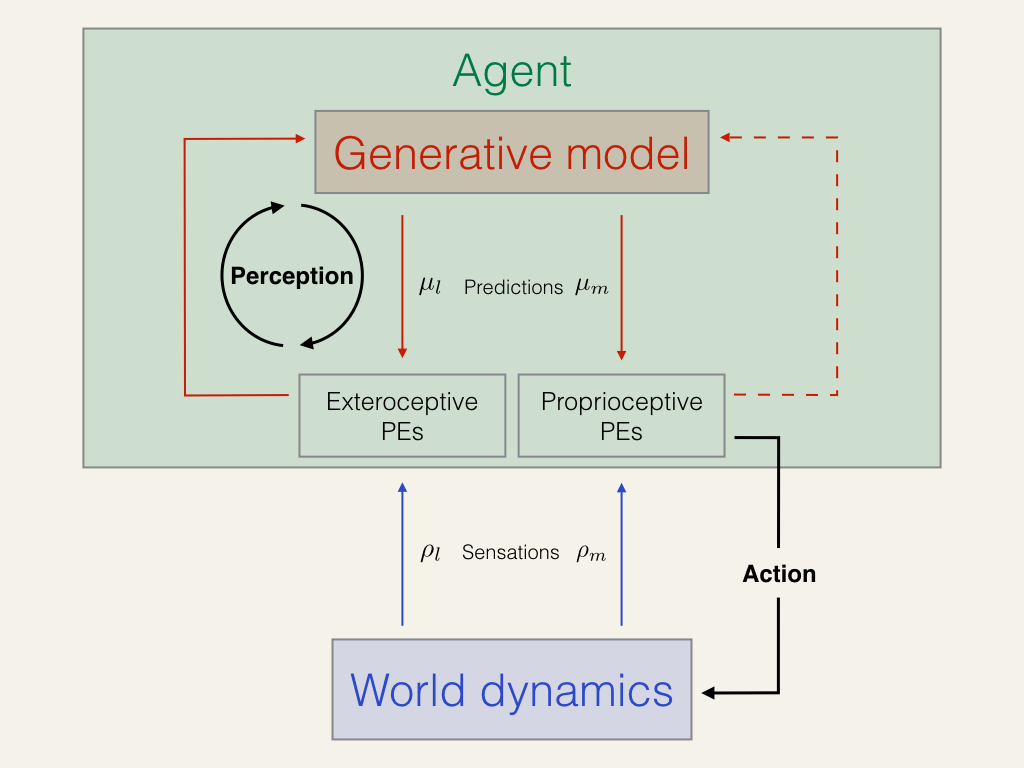}
\caption{}
\label{fig:diagram}
\end{subfigure}
\caption{(a) The wheeled vehicle used in our simulations. The agent receives input from two exteroceptors reading light intensity ($\rho_{l_1}, \rho_{l_2}$) and two proprioceptors reading wheel velocity ($\rho_{m_1}, \rho_{m_2}$). Variables $\mu_{l_1}, \mu_{l_2}, \mu_{m_1}, \mu_{m_2}$ are part of the generative model of the agent. The red lines represent the relations between the agent's prior beliefs on the dynamics of the world, very distant from how the real dynamics work. (b) A schematic of the FEP. Two types of sensations, exteroceptive (light intensity, $\rho_l$) and proprioceptive (motor velocity, $\rho_m$), represent the sensory input of the agent (blue arrows). Beliefs on causes $\mu_l$ and $\mu_m$ are updated (red arrows) within the generative model through perception. The dashed red arrow denotes the lack of update of the generative model due to proprioceptive prediction errors, necessary for phototaxis. This update is introduced later on to show ``pathological behaviour". Action $a$ solves the discrepancy between predictions of the generative model and sensations from the world by engaging with the latter.}
\end{figure}

To present some of the core ideas behind the FEP and active inference we implement phototaxis on a simple wheeled vehicle. We simulate an agent with circular body, 2 noisy light sensors and 2 noiseless motors, see \figref{fig:vehicle}. For simplicity we do not simulate occlusion of the light source by the agent's body.

In previous agent-based simulations of the FEP it is typically assumed that an agent possesses a rich and detailed model of its environment (see for example \cite{Friston2010biocyb}). For instance if we were to take this approach here, we would perhaps start by assuming that the agent has a representation of the locations of both itself and the light source. However a more action-oriented interpretation of the FEP suggests that adaptive behaviour could emerge from generative models that are more frugal and parsimonious \citep{seth2014cybernetic, clark2015radical, bruineberg2016anticipating, allen2016cognitivism}. To examine this, we endow our agent with a minimal model of its surrounding environment. Specifically, our agent receives four inputs: two from exteroceptors sensitive to light $\rho_{l_1}, \rho_{l_2}$ and two from proprioceptors  $\rho_{m_1}, \rho_{m_2}$ sensing motor velocity, see \figref{fig:vehicle}. We then assume that it only models four hidden causes $x = \{l_1, l_2, m_1, m_2\}$, one for each input, parametrised by beliefs $\mu_x = \{\mu_{l_1}, \mu_{l_2}, \mu_{m_1}, \mu_{m_2}\}$. In \tabref{tab:variables} we list the variables used in our model.

\begin{table}[h]
\centering
\def\arraystretch{1.5}
\begin{tabular}{ | c | m{13em} | } 
\hline
Variable & Meaning \\ 
\hline
$\rho$ & Set of sensory inputs $\{\rho_{l_1}, \rho_{l_2}, \rho_{m_1}, \rho_{m_2}\}$ \\
$\rho_{l_1}, \rho_{l_2}$ & Readings of luminance from sensors 1 and 2 (exteroceptors) \\ 
$\rho_{m_1}, \rho_{m_2}$ & Readings of velocity from motors 1 and 2 (proprioceptors) \\
$\mu_x$ & Set of parametrised beliefs on causes $x$ $\{\mu_{l_1}, \mu_{l_2}, \mu_{m_1}, \mu_{m_2}\}$ \\
$\mu_{l_1}, \mu_{l_2}$ & (Parametrised) Beliefs about exteroceptive sensory readings \\
$\mu_{m_1}, \mu_{m_2}$ & (Parametrised) Beliefs about proprioceptive sensory readings \\
$z, w$ & Gaussian noise representing uncertainty of the agent on sensory input and beliefs about the input, respectively \\
\hline
\end{tabular}
\caption{Variables used in our definition of the generative model.}
\label{tab:variables}
\end{table}
To specify the agent's generative density (\eqnref{eq:freeEnergyLaplace}) $P(\rho, \mu_x) = P(\rho| \mu_x) P(\mu_x)$ we must first introduce a likelihood, $P(\rho| \mu_x)$ and a prior $P(\mu_x)$ in terms of the agent's beliefs $\mu_x$. In order to do so we first define a model of how exteroceptive sensations (light intensity) are generated according to the agent:
\begin{eqnarray}
\rho_{l_1} = \mu_{l_1} + z_{l_1}, \quad \rho_{l_2} = \mu_{l_2} + z_{l_2},
\label{eq:likelihoodExtero}
\end{eqnarray}
and similarly for the proprioceptors, representing readings of the velocity of each motor:
\begin{eqnarray}
\rho_{m_1} = \mu_{m_1} + z_{m_1}, \quad \rho_{m_2} = \mu_{m_2} + z_{m_2}
\label{eq:likelihoodProprio}
\end{eqnarray}
where we have assumed sensory reading are linearly related to their causes, with some additive zero-mean Gaussian noise $z = \{z_{l_1}, z_{l_2}, z_{m_1}, z_{m_2}\}$ with variance $\sigma^2_{z} = \{\sigma^2_{z_{l_1}}, \sigma^2_{z_{l_2}}, \sigma^2_{z_{m_1}}, \sigma^2_{z_{m_2}}\}$. The agent's priors on hidden causes are then specified in terms of the relation $P(\mu_m, \mu_l) = P(\mu_m | \mu_l) P(\mu_l)$, with variables $\mu_{m_1}, \mu_{m_2}$ only depending on $\mu_{l_1}, \mu_{l_2}$. We then write a model of the priors as:
\begin{eqnarray}
\mu_{m_1} = \mu_{l_2} + w_{m_1}, \quad \mu_{m_2} = \mu_{l_1} + w_{m_2}
\label{eq:prior}
\end{eqnarray}
where $w = \{w_{m_1}, w_{m_2}\}$ is some zero-mean Gaussian noise with variance $\sigma^2_{w} = \{\sigma^2_{w_{m_1}}, \sigma^2_{w_{m_2}}\}$. Effectively, we describe the underlying dynamics in terms of a contralateral relationship between beliefs about sensors $\mu_l$ and motors $\mu_m$. As we will see, this beliefs' structure makes our agent functionally consistent with Braitenberg vehicle 2b, the ``aggressor" \citep{braitenberg1986vehicles}. We also assume uniform priors on beliefs about exteroceptors $P(\mu_l)$, thus eliminating them from our formulation.

Under the assumption that random variables $z$ are Gaussian with zero mean, $\mathcal{N}(0, \sigma^2) = 1 / \sqrt{2\pi \sigma^2} \exp{(-z^2/(2\sigma^2))}$, by rewriting them as
\begin{eqnarray}
& z_{l_1} = \rho_{l_1} - \mu_{l_1}, \quad z_{l_2} = \rho_{l_2} - \mu_{l_2} \\
& z_{m_1} = \rho_{m_1} - \mu_{m_1}, \quad z_{m_2} = \rho_{m_2} - \mu_{m_2}
\end{eqnarray}
we can define the likelihood functions as
\begin{eqnarray}
& P(\rho_l | \mu_l) = \frac{1}{\sqrt{2\pi \sigma_{z_l}^2}} \exp\Big({\frac{-(\rho_l - \mu_l)^2}{(2\sigma_{z_l}^2)}}\Big) \nonumber \\
& P(\rho_m | \mu_m) = \frac{1}{\sqrt{2\pi \sigma_{z_m}^2}} \exp\Big({\frac{-(\rho_m - \mu_m)^2}{(2\sigma_{z_m}^2)}}\Big)
\end{eqnarray}
where $l = \{l_1, l_2\}$ and $m = \{m_1, m_2\}$. Similarly, with Gaussian noise $w$ the priors become
\begin{eqnarray}
& P(\mu_{m_1} | \mu_{l_2}}) = \frac{1}{\sqrt{2\pi \sigma_{w_{m_1}}^2}} \exp\Big({\frac{-(\mu_{m_1} - \mu_{l_2})^2}{(2\sigma_{w_{m_1}}^2)}\Big) \nonumber \\
& P(\mu_{m_2} | \mu_{l_1}}) = \frac{1}{\sqrt{2\pi \sigma_{w_{m_2}}^2}} \exp\Big({\frac{-(\mu_{m_2} - \mu_{l_1})^2}{(2\sigma_{w_{m_2}}^2)}\Big)
\end{eqnarray}

As a result of the assumption for both $z$ and $w$ to be Gaussian, the free energy reduces to (without any constant):
\begin{eqnarray}
& F = \frac{1}{2} \Big( \sum_i \pi_{z_i} (\rho_i - \mu_i)^2 + \pi_{w_{m_1}} (\mu_{m_1} - \mu_{l_2})^2 \nonumber + \\ & + \pi_{w_{m_2}} (\mu_{m_2} - \mu_{x_1})^2 + \sum_i \ln \pi_{z_i} + \sum_j \ln \pi_{w_j} \Big)
\label{eq:LaplaceFE}
\end{eqnarray}
with $i \in \{l_1, l_2, m_1, m_2\}$ and $j \in \{m_1, m_2\}$ and where $\pi_{z_i}$ and $\pi_{w_j}$ are the inverse variances of noise terms $z_i$ and $w_j$ respectively, also called ``precisions". Precision parameters weight predictions errors based on the agent's confidence on a certain belief. High precisions imply low variances and thus high confidence, and vice versa. These parameters allow for different emphases on the minimisation of free energy, for example an agent could focus more on predictions weighted by $\pi_{w_j}$ or rely more on sensations from the environment when $\pi_{z_i}$ are large. Some of the implications of this weighting mechanism will be developed in more detail for this model with our simulations.

With the expression for the free energy we can now derive the equations that will implement perception (\eqnref{eq:perceptionMin}):
\begin{eqnarray}
&\dot{\mu}_{l_1} = - k \Big( \pi_{z_{l_1}} (\mu_{l_1} - \rho_{l_1}) + \pi_{w_{m_2}} (\mu_{l_1} - \mu_{m_2}) \Big) \nonumber \\
&\dot{\mu}_{l_2} = - k \Big( \pi_{z_{l_2}} (\mu_{l_2} - \rho_{l_2}) + \pi_{w_{m_1}} (\mu_{l_2} - \mu_{m_1}) \Big) \nonumber \\
&\dot{\mu}_{m_1} = - k \Big( \pi_{z_{m_1}} (\mu_{m_1} - \rho_{m_1}) + \pi_{w_{m_1}} (\mu_{m_1} - \mu_{l_2}) \Big) \nonumber \\
&\dot{\mu}_{m_2} = - k \Big( \pi_{z_{m_1}} (\mu_{m_2} - \rho_{m_2}) + \pi_{w_{m_2}} (\mu_{m_2} - \mu_{l_1}) \Big)
\label{eq:perception}
\end{eqnarray}
and action (\eqnref{eq:actionMin}):
\begin{eqnarray}
\dot{a}_1 = & - k \Big( \pi_{z_{m_1}} (\rho_{m_1} - \mu_{m_1}) \frac{\partial \rho_{m_1}}{\partial a_1} + \nonumber \\
& + \pi_{z_{m_2}} (\rho_{m_2} - \mu_{m_2}) \frac{\partial \rho_{m_2}}{\partial a_1} \Big) \nonumber \\
\dot{a}_2 = &- k \Big( \pi_{z_{m_1}} (\rho_{m_1} - \mu_{m_1}) \frac{\partial \rho_{m_1}}{\partial a_2} + \nonumber \\
& + \pi_{z_{m_2}} (\rho_{m_2} - \mu_{m_2}) \frac{\partial \rho_{m_2}}{\partial a_2} \Big)
\label{eq:action}
\end{eqnarray}
where $k$ is the learning rate used for the gradient descent.

To implement action according to \eqnref{eq:actionMin} we must first define the partial derivative $\partial \rho/\partial a$ to implement the agent's model of the relationship between actions and percepts. Here we assume that actions ($a_1, a_2$) can only influence proprioceptive sensations $\rho_{m_1}, \rho_{m_2}$. As explained in \cite{Friston2010biocyb}, active inference dispenses the traditional notion of forward/inverse models for motor control in favour of a more general generative (forward) model inverted through Bayesian inference. In this framework, inverse models are thought to be implicitly encoded in predictions about proprioceptive consequences that are implemented through simple reflex arcs thought to be embodied in an agent. In our model the inverse model corresponds to the partial derivatives $\partial \rho/\partial a$'s, encoded as
\begin{eqnarray}
\begin{bmatrix}
\frac{\partial \rho_{m_1}}{\partial a_1} & \frac{\partial \rho_{m_2}}{\partial a_1} \\
\frac{\partial \rho_{m_1}}{\partial a_2} & \frac{\partial \rho_{m_2}}{\partial a_2}
\end{bmatrix} = 
\begin{bmatrix}
1 & 0 \\
0 & 1
\end{bmatrix}
\end{eqnarray}
to specify at a proprioceptive level cross relations between sensors and motors. Motor reflex arcs are implemented with the assumption that each motor takes as an input one of the actions produced through the minimisation of free energy
\begin{eqnarray}
v_1 = a_1, \quad v_2 = a_2.
\end{eqnarray}
where $v_1, v_2$ are the actual motor velocities of the vehicle.

\section{Simulations}

\subsection{Phototaxis}

\begin{figure*}[ht!]
\begin{subfigure}{.32\textwidth}
  \centering
  \includegraphics[width=\linewidth]{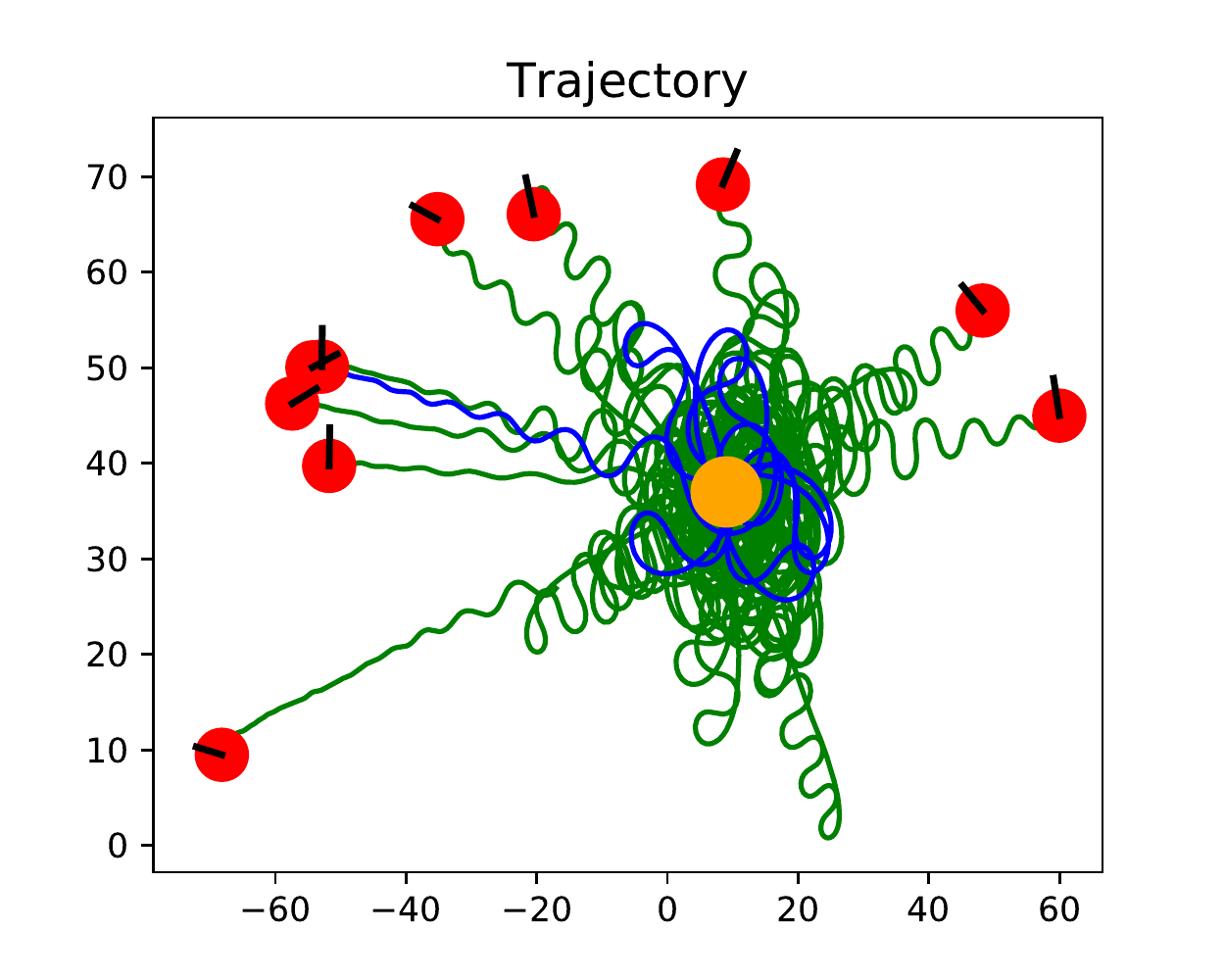}
  \caption{}
  \label{fig:TrajectoryPhototaxis}
\end{subfigure}
\begin{subfigure}{.32\textwidth}
  \centering
  \includegraphics[width=\linewidth]{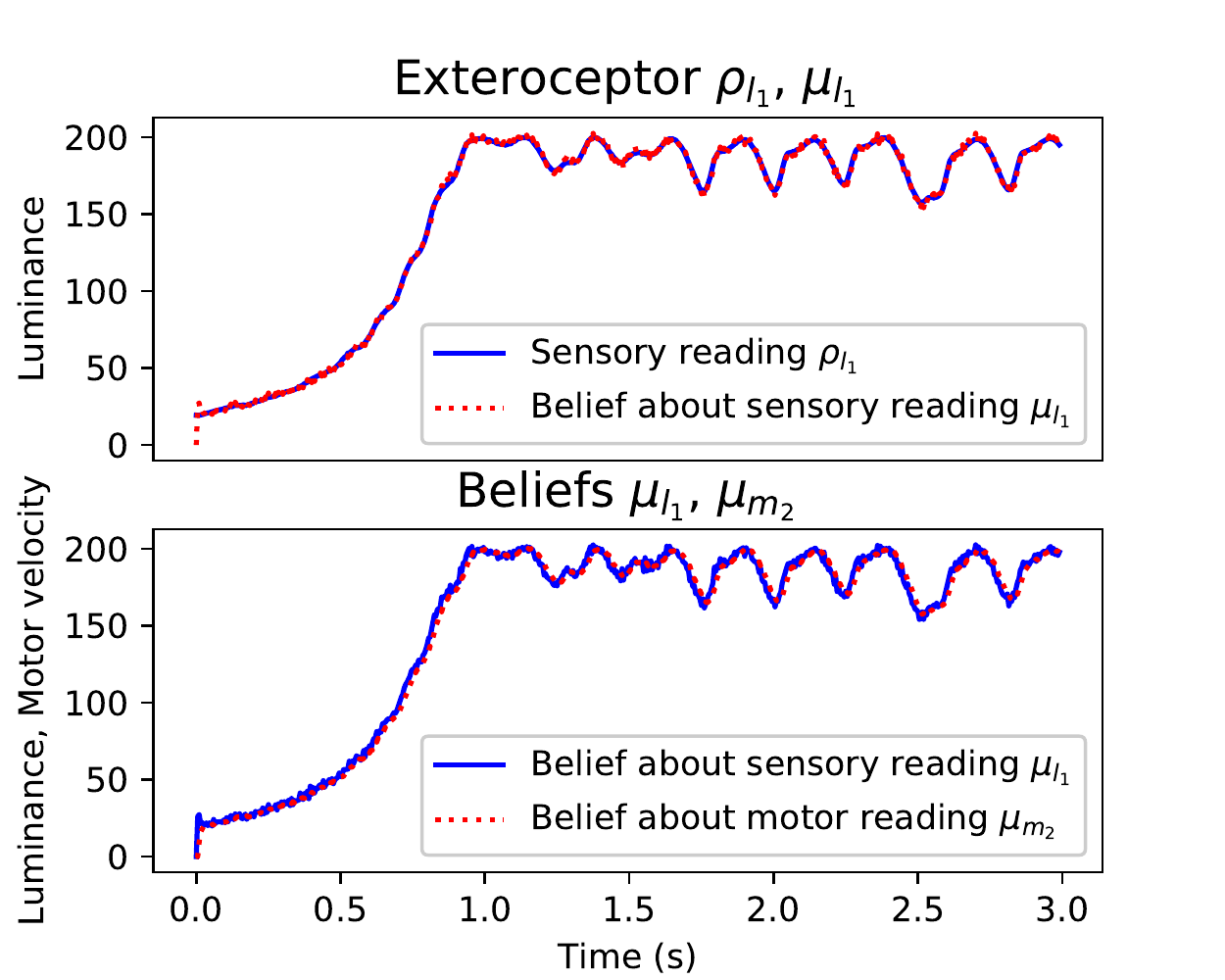}
  \caption{}
  \label{fig:ExteroceptorBeliefsPhototaxis}
\end{subfigure}
\begin{subfigure}{.32\textwidth}
  \centering
  \includegraphics[width=\linewidth]{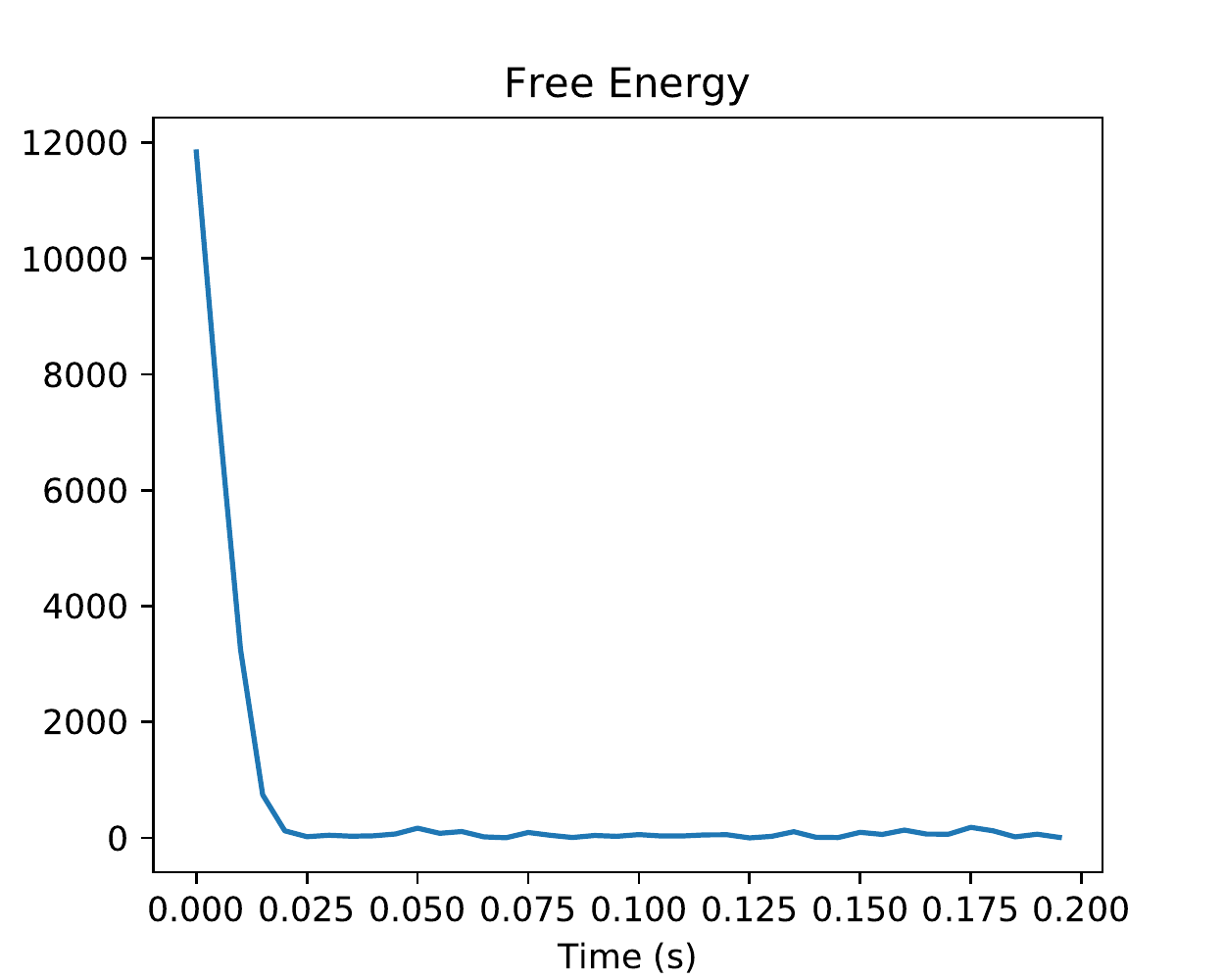}
  \caption{}
  \label{fig:FEPhototaxis}
\end{subfigure}

\begin{subfigure}{.32\textwidth}
  \centering
  \includegraphics[width=\linewidth]{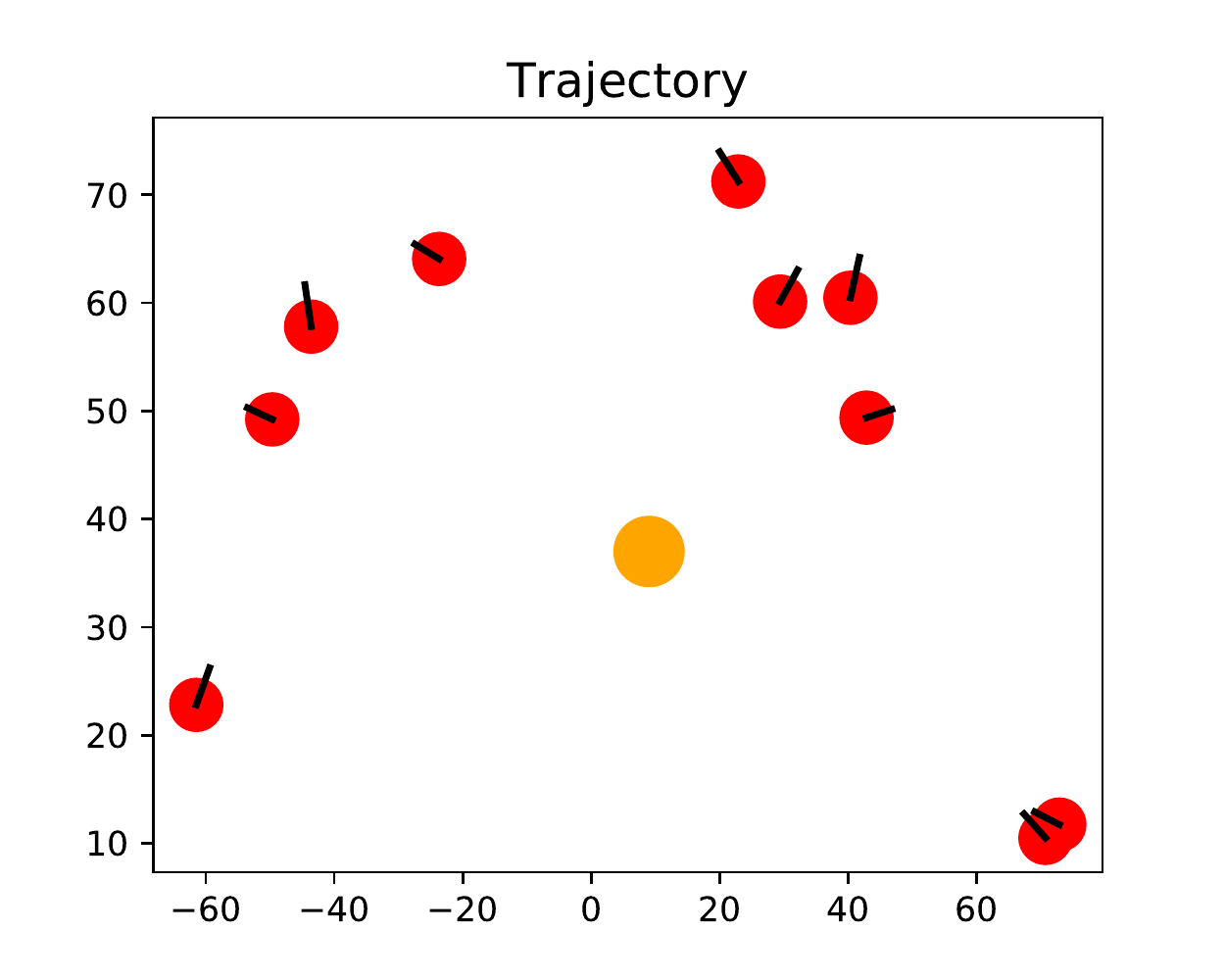}
  \caption{}
  \label{fig:TrajectoryHighPiZPro}
\end{subfigure}
\begin{subfigure}{.32\textwidth}
  \centering
  \includegraphics[width=\linewidth]{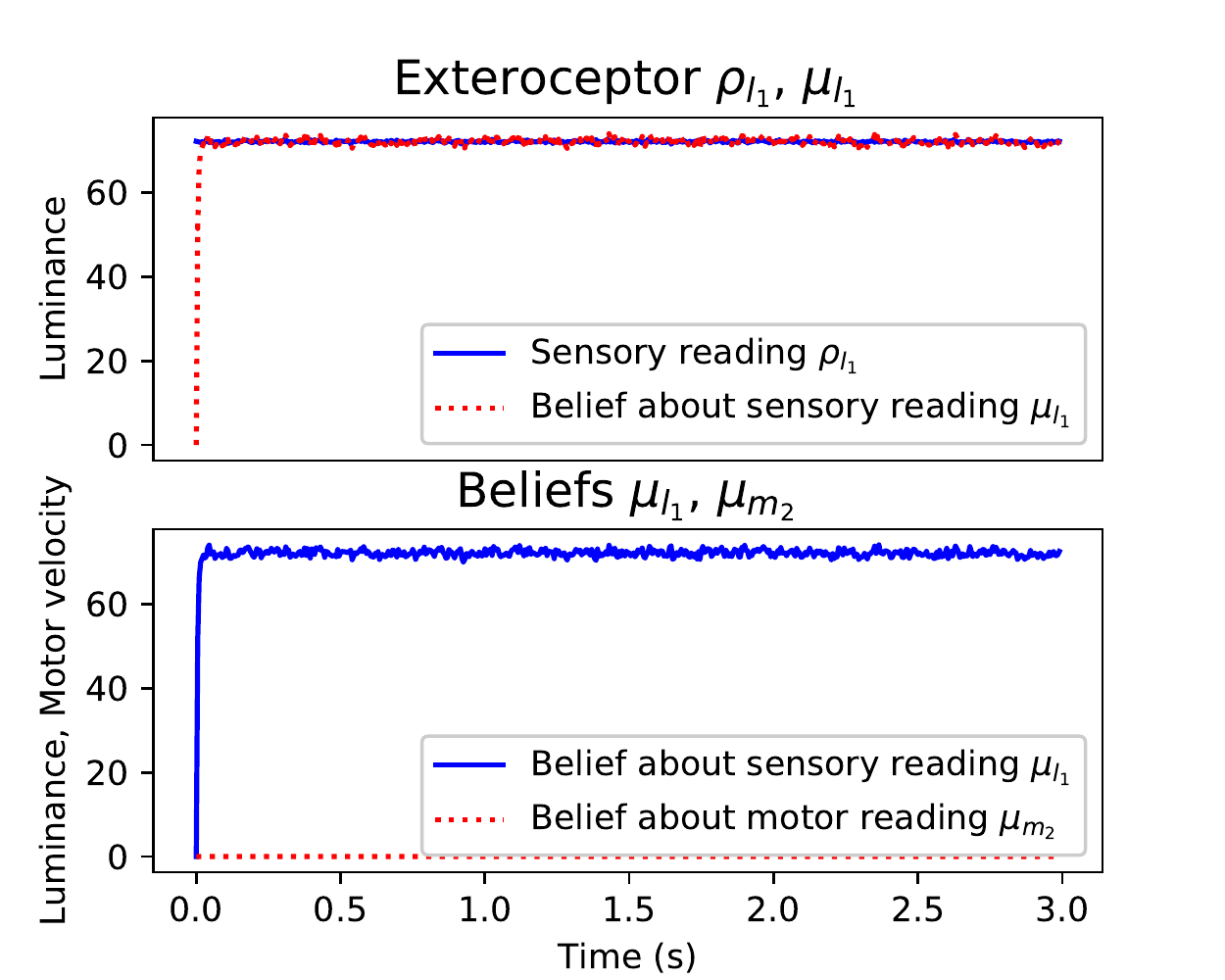}
  \caption{}
  \label{fig:ExteroceptorBeliefsHighPiZPro}
\end{subfigure}
\begin{subfigure}{.32\textwidth}
  \centering
  \includegraphics[width=\linewidth]{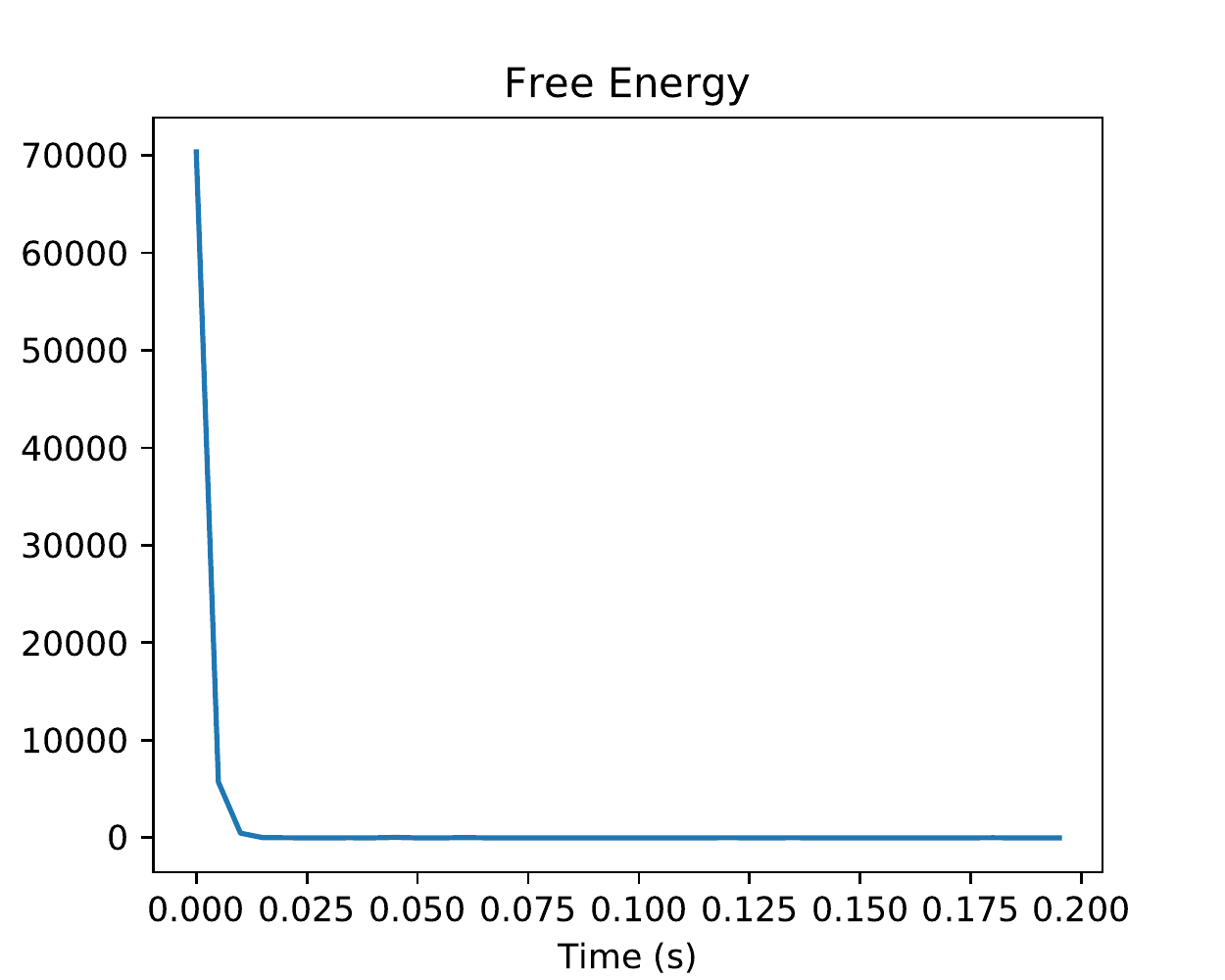}
  \caption{}
  \label{fig:FEHighPiZPro}
\end{subfigure}
\caption{First row: Phototaxis under the FEP/Active Inference. (a) Trajectory of the vehicle over 5 seconds of simulation. The red dot represents the initial position (with orientation parallel to the vertical axis), the yellow circle is the light source. (b) Inference of belief $\mu_{l_1}$ about sensory input $\rho_{l_1}$. (c) Coupling of belief on left sensor $\mu_{l_1}$ with belief on right motor $\mu_{m_2}$. (d) Free energy over the first 0.2 seconds. The vehicle performs phototaxis akin to Braitenberg vehicle 2b, by accelerating close to the light. However the free energy (only shown for the agent describing the blue trajectory) is minimised long before ($\approx 0.05s$) this agent gets close to the light source ($\approx 2s$), since the generative model only encodes conditions on sensory-motor coupling (e.g. no specification of a light level as a target). The only requirement for this agent is to fulfill the coupling between sensors and motors specified by its generative model.\\ \\
Second row: Example of ``pathological" behaviour under the FEP/Active Inference (akinesia). Same layout as above. Here movements are inhibited (\figref{fig:TrajectoryHighPiZPro}) by high precisions on proprioceptive prediction errors and low precisions on the priors. This prevents the appropriate coupling between beliefs about sensors and motors necessary for phototaxis (\figref{fig:ExteroceptorBeliefsHighPiZPro} bottom). \\ \\
Each row shows the simulation of 10 different vehicles with random initial conditions (position and orientation of the agent). We also added Gaussian noise to each precision parameter (mean varying up to $\pm 20\%$ of the value of each precision parameter, variance 1) to confirm the robustness of the solution.}
\label{fig:PhototaxisAkinesia}
\end{figure*}

The behaviour of agents operating under the FEP can be shown to depend on the value of precision parameters. In our first simulation we adjust the precisions to implement phototaxis. Priors in the generative model implement linear contralateral relations between beliefs $\mu_l$ and $\mu_m$ (see \eqnref{eq:prior}). In this setup, our agent needs to accurately infer $\mu_l$ from its readings on luminance $\rho_l$ (\figref{fig:ExteroceptorBeliefsPhototaxis} top). Beliefs $\mu_l$ are then mapped to beliefs about proprioceptive input $\mu_m$ (\figref{fig:ExteroceptorBeliefsPhototaxis} bottom) and projected to the motors via fast (instantaneous in our simulations) reflex arcs, i.e. $\rho_m = \mu_m$. To do so, precisions on proprioceptive sensations ($\pi_{z_m}$) need to be much lower relatively to those on the priors ($\pi_{w_m}$), in turn smaller than the precisions on exteroceptors ($\pi_{z_l}$).

This enables the sensory-motor flow we described: 1) inference of exteroceptive sensations, 2) coupling between beliefs about light levels and motor velocities and 3) actuation of the motors based on the agent's beliefs about proprioceptive input (dictated by the light intensity). In \tabref{tab:behavioursEq} we show how the the formulation of perception and action set out in \eqnref{eq:perception} and \eqnref{eq:action} respectively is simplified by these assumptions in the left-sensor/right-motor relation, $\{\mu_{l_1}, \mu_{m_2}\}$ (the same goes for $\{\mu_{l_2}, \mu_{m_1}\}$).

\figref{fig:TrajectoryPhototaxis} shows the ``aggressor-like" behaviour of our agent, which speeds up close to the light and slows down away from it. At the beginning of our simulation we initialise all beliefs to zero. \figref{fig:FEPhototaxis} shows how, after a brief transient due to these initial conditions, the free energy rapidly approaches a minimum value exhibiting fluctuations only driven by noise on exteroceptive input. This minimum is reached in less than 0.1 seconds even though it takes the agent $\approx 2$ seconds to reach the light, see \figref{fig:ExteroceptorBeliefsPhototaxis} top. This is because the generative model we define does not encode explicit priors on light levels and thus does not specify a target luminance. Instead the agent minimises free energy by simply satisfying a mapping between beliefs about light levels and motor velocities. Phototaxis emerges as a consequence of the relationships between beliefs $\mu_l$ and $\mu_m$, not because of an explicitly encoded goal. We will come back to this idea in the discussion.

\subsection{Pathological behaviour}
The sensitivity of behaviour to precisions has been used extensively to develop hypotheses and computational models for phenomena including for instance psychosis and schizophrenia \citep{adams2013psychosis, friston2016dysconnection}, sensory attenuation \citep{brown2013active} and attention \citep{feldman2010attention}. Here we present an interpretation of the role of precisions for our agent inspired by these accounts.


If the balance of precisions on proprioceptive inputs $\pi_{z_m}$ and on priors $\pi_{w_m}$ is altered, with an increase of the former and a decrease of the latter, the behaviour of our vehicle is severely affected by the change. The sensory-motor flow necessary for phototaxis is disrupted and while the agent still infers light levels through $\mu_l$ (\figref{fig:ExteroceptorBeliefsHighPiZPro} top), these beliefs are not mapped to $\mu_m$ (\figref{fig:ExteroceptorBeliefsHighPiZPro} bottom) since the agent's precisions about their relation $\pi_{w_m}$ are dominated by precisions on proprioceptive input $\pi_{z_m}$ (equation II in \tabref{tab:behavioursEq}).

This behaviour is consistent with interpretations under the FEP of motor control disorders where movements are limited (hypokinesia) or entirely absent (akinesia) \citep{brown2013active}. In active inference terms, decreasing proprioceptive prediction errors precisions, $\pi_{z_m}$ in our case, is thought to be a necessary condition to actuate motor commands \citep{brown2013active}. Failing to reduce them generates atypical behaviour corresponding to an agent completely unable to move as in the case we discussed when precisions on proprioceptive input $\pi_{z_m}$ are larger than precisions $\pi_{w_m}$ (\figref{fig:TrajectoryHighPiZPro}), or having limited motor capabilities when $\pi_{z_m}$ and $\pi_{w_m}$ are closer in magnitude (not shown here).

\begin{table}[h]
\centering
\def\arraystretch{1.5}
\begin{tabular}{ | m{13em} |} 
\hline
Phototaxis \newline I. $\dot{\mu}_{l_1} \approx - \pi_{z_{l_1}} (\mu_{l_1} - \rho_{l_1})$ \newline II. $\dot{\mu}_{m_2} \approx - \pi_{w_{m_2}} (\mu_{m_2} - \mu_{l_1})$ \newline III. $\dot{a}_2 \approx - \pi_{z_{m_2}} (\rho_{m_2} - \mu_{m_2})$ \\ 
Pathological \newline I. $\dot{\mu}_{l_1} \approx - \pi_{z_{l_1}} (\mu_{l_1} - \rho_{l_1})$ \newline II. $\dot{\mu}_{m_2} \approx - \pi_{z_{m_2}} (\mu_{m_2} - \rho_{m_2})$ \newline III. $\dot{a}_2 \approx - \pi_{z_{m_2}} (\rho_{m_2} - \mu_{m_2})$ \\ 
\hline
\end{tabular}
\caption{Approximations of \eqnref{eq:perception} and \eqnref{eq:action} after taking different assumptions on precisions. Equations for both phototaxis and a possible pathological behaviour are presented for the left-sensor/right-motor coupling.}
\label{tab:behavioursEq}
\end{table}

\subsection{Other vehicles}
In the previous sections we showed that different precisions within the same generative model can qualitatively affect the behaviour of an agent, going from performing phototaxis to a catatonic state by just regulating some of these weights. In this section we explore how different priors, encoding in our case the relationships between exteroceptive and proprioceptive inputs (see \eqnref{eq:prior}) can also be used to generate new emergent behaviour. As we saw in our ``Model" section during the definition of our generative model, vehicles in \figref{fig:TrajectoryPhototaxis} described trajectories consistent with Braitenberg vehicle 2b. Simple modifications to the priors can easily reproduce behaviours analogous to other vehicles, for instance 2a (the ``coward"), 3a (the ``lover") and 3b (the ``explorer").

In more detail, models of the priors were updated to:
\begin{itemize}
    \item for the coward vehicle, rapidly running away from the light to then slow down once distance is gained, \begin{eqnarray}
        \mu_{m_1} = \mu_{l_1} + w_{m_1}, \quad \mu_{m_2} = \mu_{l_2} + w_{m_2}
        \label{eq:priorCoward}
    \end{eqnarray}
    \item for the lover vehicle, quickly approaching the light while decreasing its speed close to the source, \begin{eqnarray}
        \mu_{m_1} = l_{max} - \mu_{l_1} + w_{m_1}, \quad \mu_{m_2} = l_{max} - \mu_{l_2} + w_{m_2}
        \label{eq:priorLover}
    \end{eqnarray}
    \item for the explorer vehicle, repelled by the light, moving slowly in its vicinity and speeding up away from it, \begin{eqnarray}
        \mu_{m_1} = l_{max} - \mu_{l_2} + w_{m_1}, \quad \mu_{m_2} = l_{max} - \mu_{l_1} + w_{m_2}
        \label{eq:priorExplorer}
    \end{eqnarray}
\end{itemize}
with $l_{max}$ representing the highest light intensity in the agents' environment.

\begin{figure}[ht!]
\begin{center}
\begin{subfigure}{.2\textwidth}
  \centering
  \includegraphics[width=\linewidth]{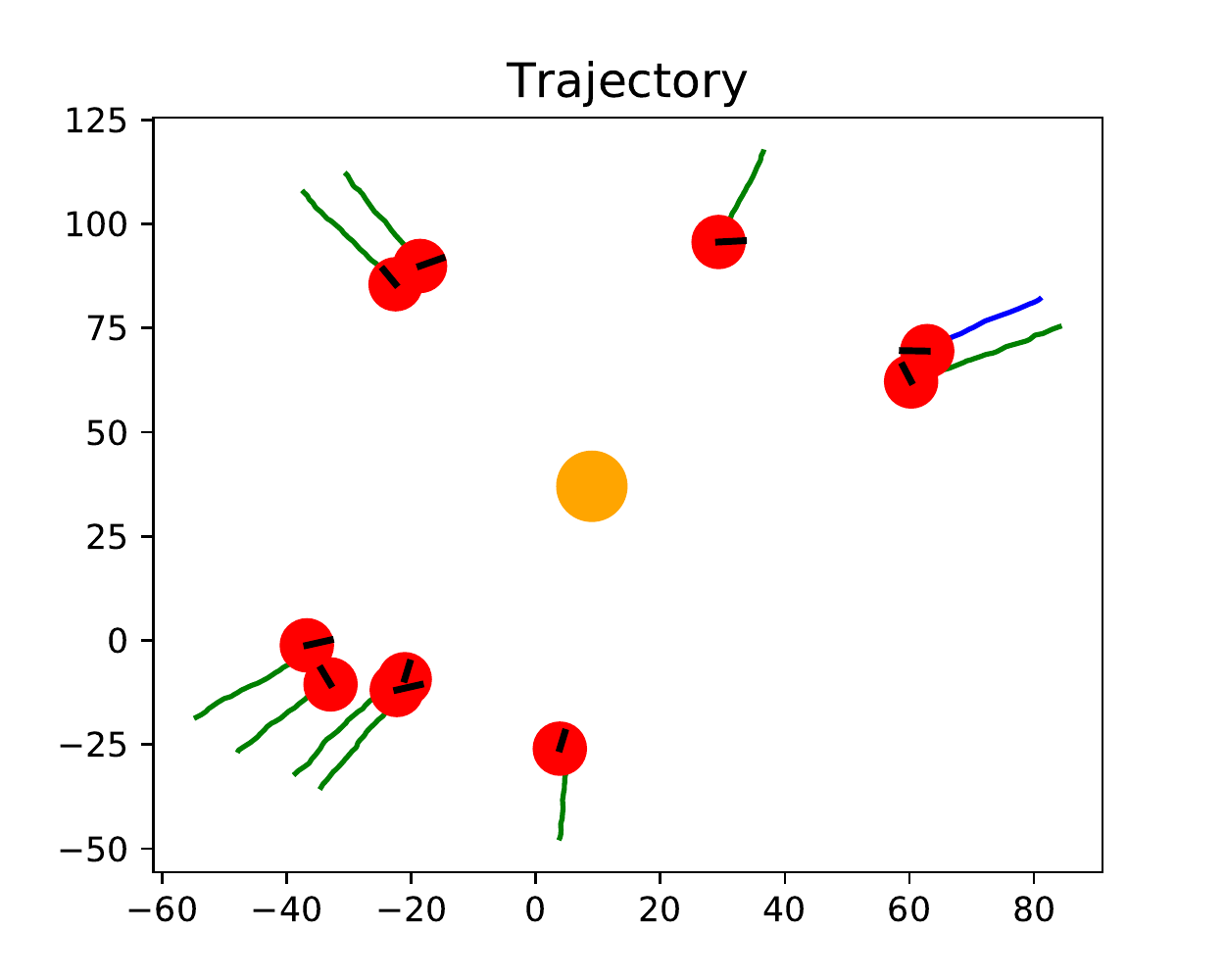}
  \caption{}
  \label{fig:TrajectoryCoward}
\end{subfigure}
\begin{subfigure}{.2\textwidth}
  \centering
  \includegraphics[width=\linewidth]{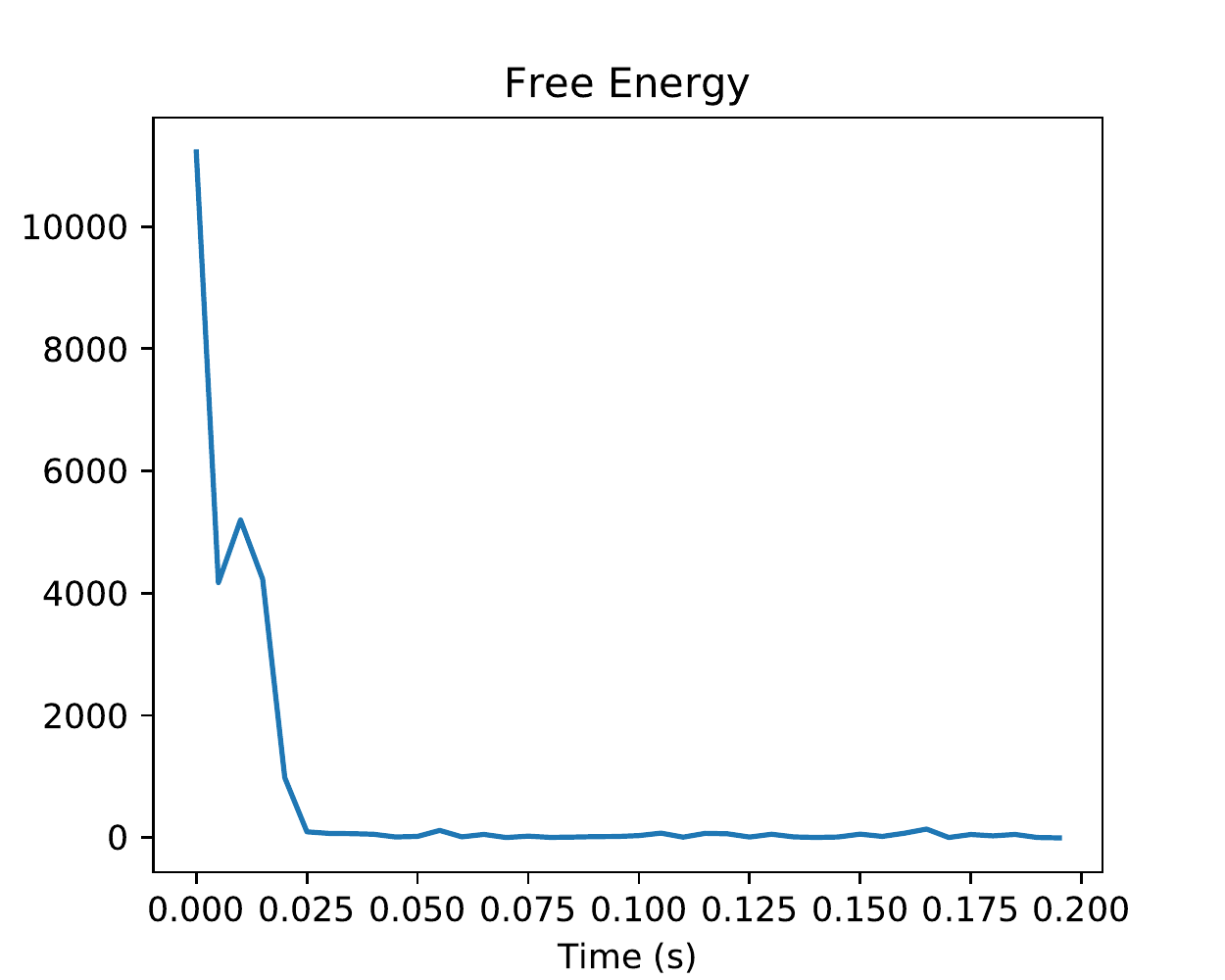}
  \caption{}
  \label{fig:FreeEnergyCoward}
\end{subfigure}
\begin{subfigure}{.2\textwidth}
  \centering
  \includegraphics[width=\linewidth]{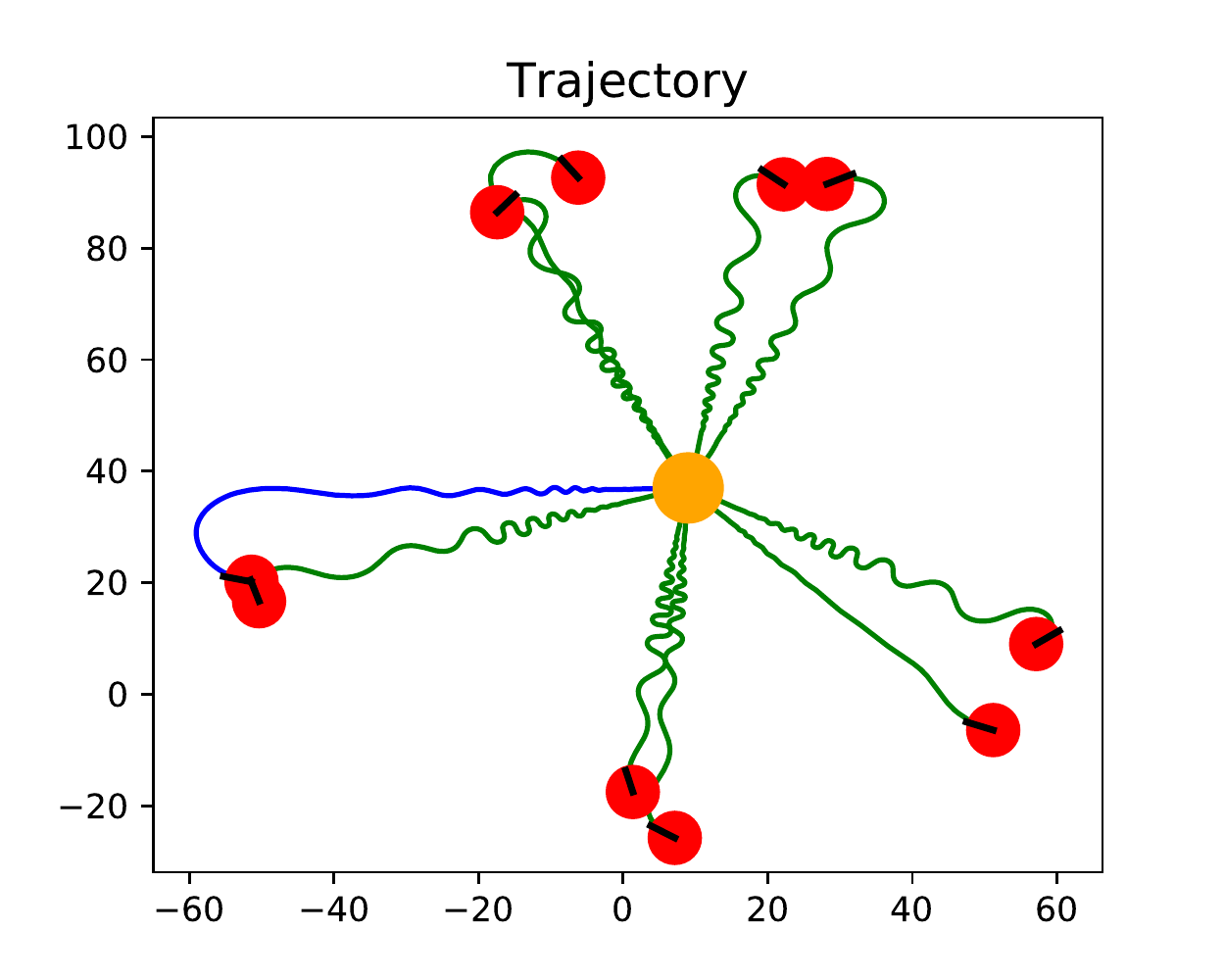}
  \caption{}
  \label{fig:TrajectoryLover}
\end{subfigure}
\begin{subfigure}{.2\textwidth}
  \centering
  \includegraphics[width=\linewidth]{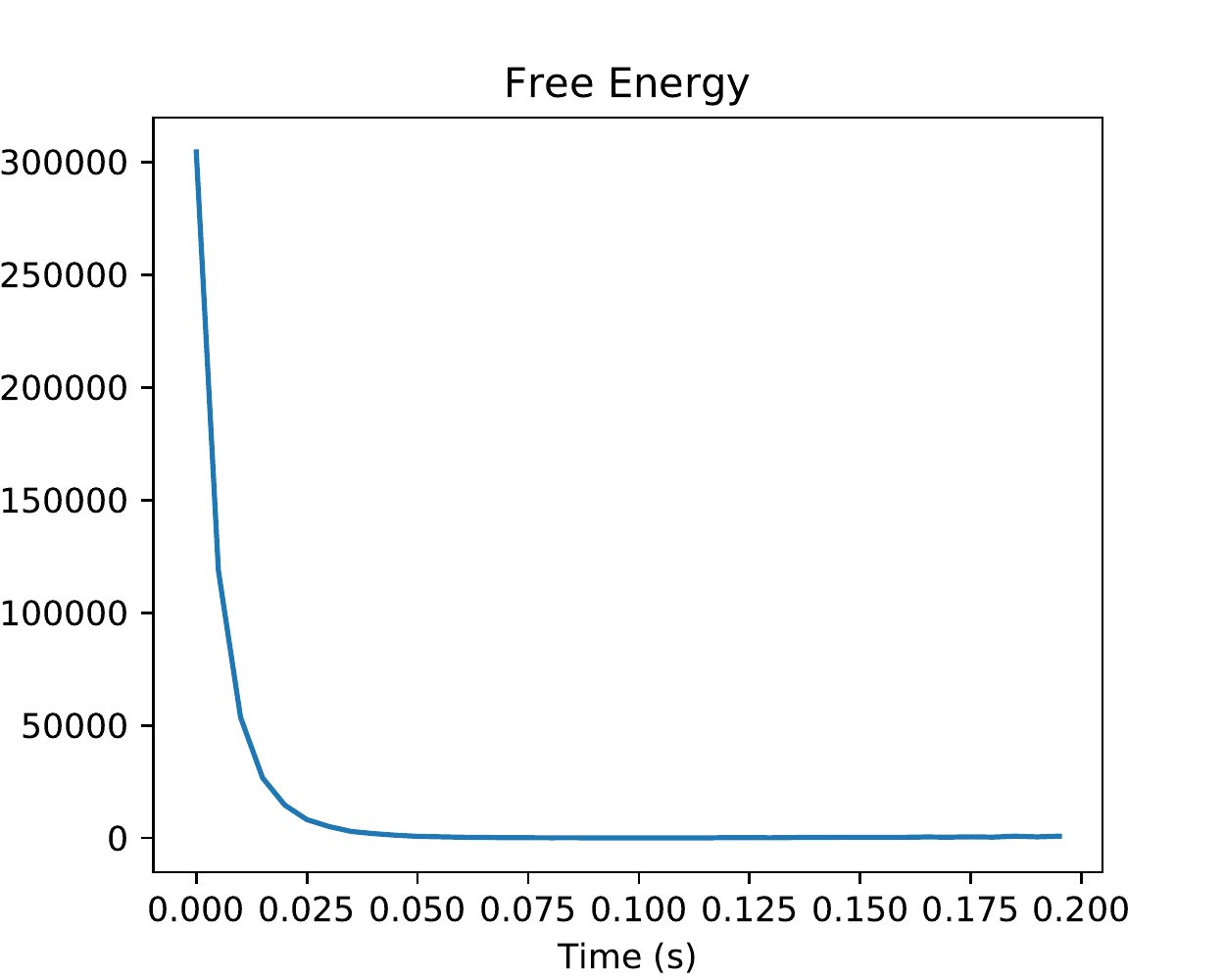}
  \caption{}
  \label{fig:FreeEnergyLover}
\end{subfigure}
\begin{subfigure}{.2\textwidth}
  \centering
  \includegraphics[width=\linewidth]{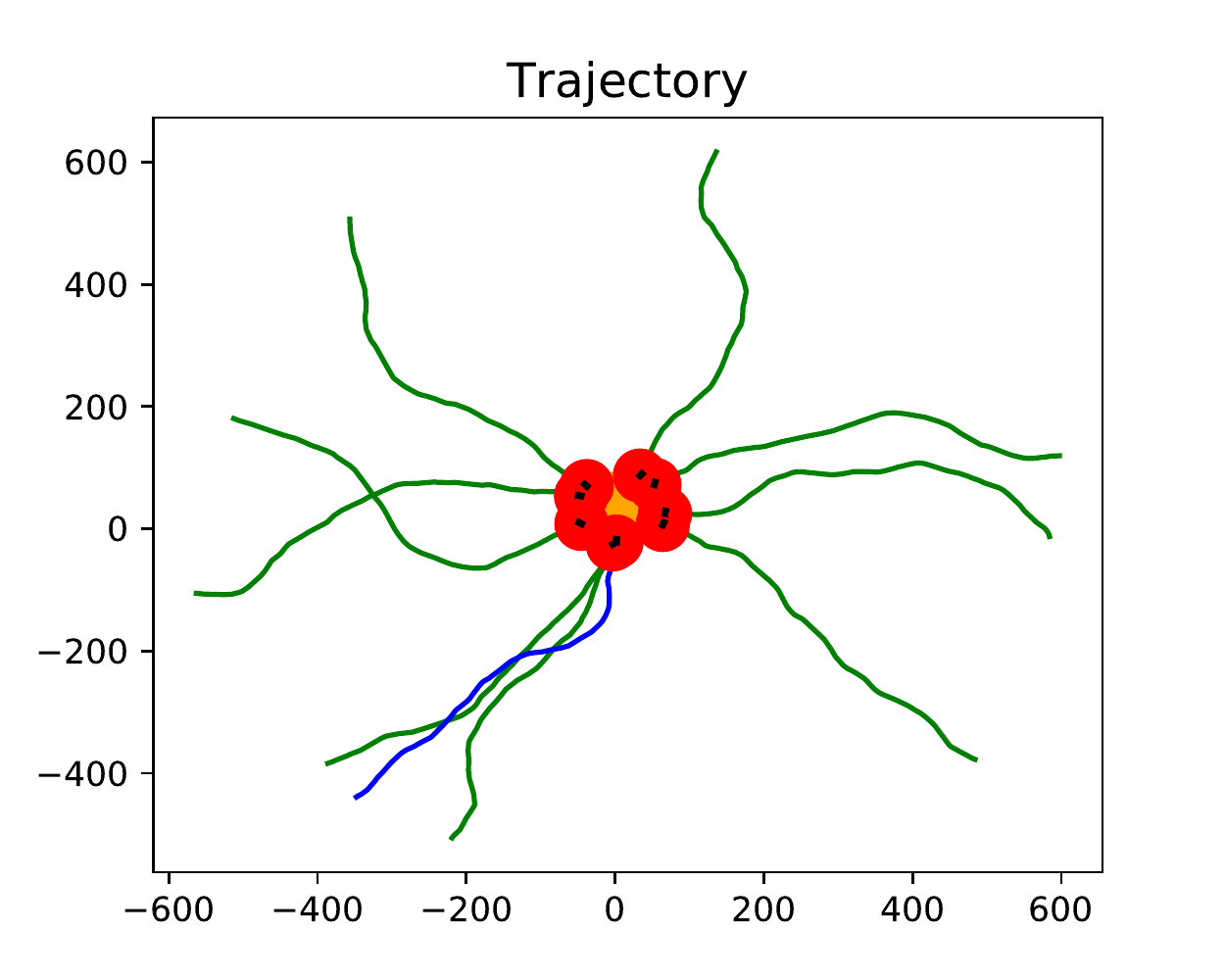}
  \caption{}
  \label{fig:TrajectoryExplorer}
\end{subfigure}
\begin{subfigure}{.2\textwidth}
  \centering
  \includegraphics[width=\linewidth]{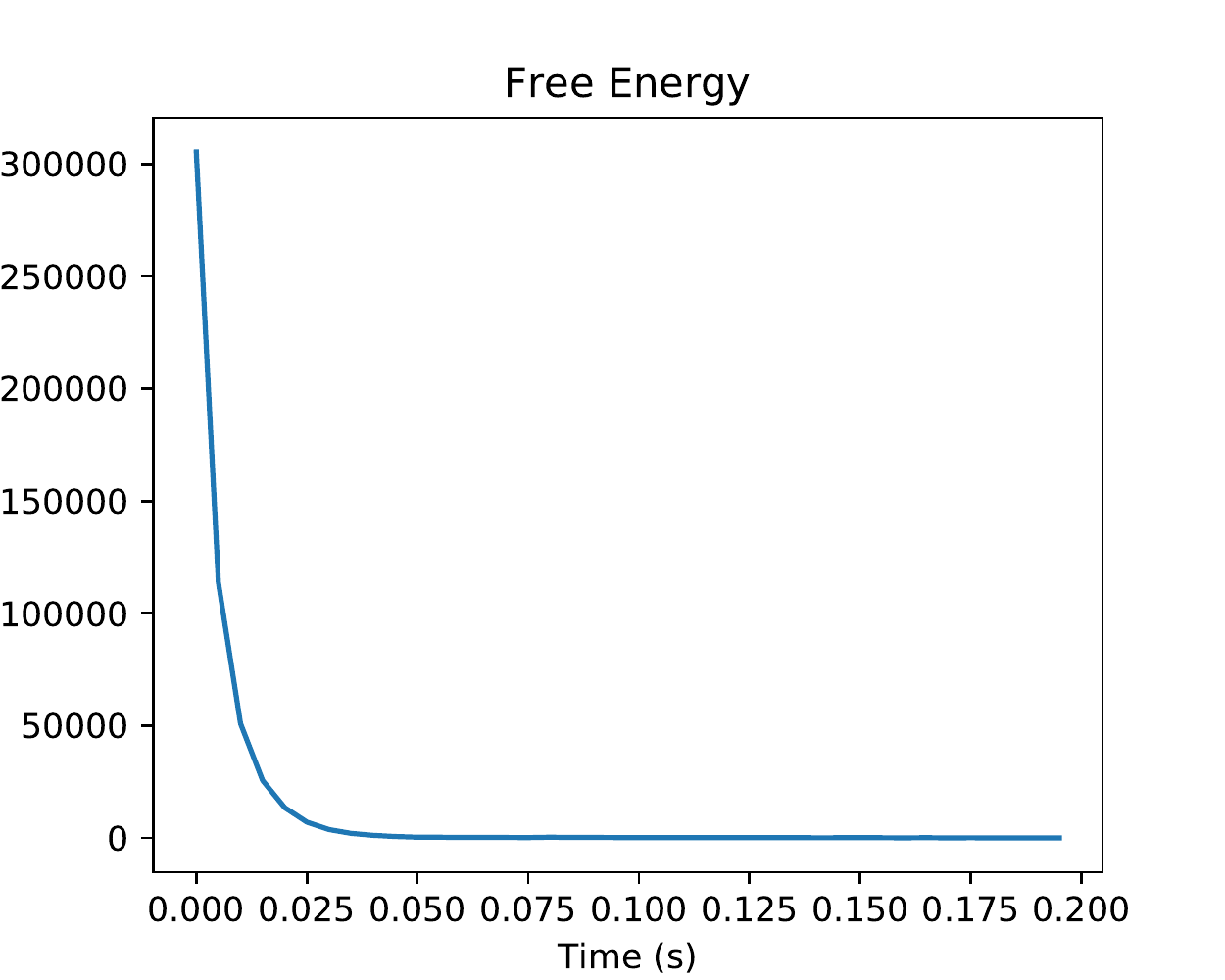}
  \caption{}
  \label{fig:FreeEnergyExplorer}
\end{subfigure}
\end{center}
\caption{Different behaviours are obtained by updating the priors of the agent. \figref{fig:TrajectoryCoward} shows the ``coward"-like behaviour of Braitenberg vehicle 2a, \figref{fig:TrajectoryLover} is akin to vehicle 3a, the ``lover", while \figref{fig:TrajectoryExplorer} behaves as vehicle 3b, the ``explorer". Ten simulations were performed for each setup, with random initial conditions (position and orientation of the agent). We also added Gaussian noise to each precision parameter (mean varying up to $\pm 20\%$ of the value of each precision parameter, variance 1) to confirm the robustness of the solution. The free energy (only shown for the agent describing the blue trajectory) is minimised in all cases as soon as the conditions described by the agents' priors are met.}
\end{figure}

\section{Discussion}
In this work we presented an implementation of phototaxis within an emerging framework in computational and cognitive neuroscience, the Free Energy Principle (FEP). According to the FEP, processes like perception, learning and action can be defined in biological systems as the minimisation of a quantity defined as (variational) free energy \citep{Friston2010nature}. Most of the simulations proposed so far have focused on ``perception-oriented" interpretations of the FEP, where generative models play the role of accurate descriptions of the dynamics of the world they represent. These models can capture the intrinsic properties of an agent's sensations and reconstruct the causes of these sensations to a great degree of detail. On this view, a model is assessed on how accurately it can predict incoming sensations, behaviour only emerges as a consequence of reliable information encoded by the agent.

Our model on the other hand, represents an example of ``action-oriented" interpretations of the FEP framework \citep{seth2014cybernetic, clark2015radical, bruineberg2016anticipating, allen2016cognitivism}, with a focus on minimal generative models. In this case, models are evaluated based on their ability to allow an agent to perform a task or achieve a certain goal. The information recapitulated in these models is only apt to perform ecologically relevant behaviour: inferring and encoding more properties of the world dynamics is not an advantage for an agent. This idea is described within the formalism of the FEP as the complexity-accuracy trade off. It is possible in fact to rewrite \eqnref{eq:freeEnergyKL} in information theoretical terms as a combination of measures of complexity and accuracy \citep{friston2006free, Friston2010nature}. According to this alternative reading, minimising free energy is equivalent to maximising the predictive power (i.e. accuracy) of a generative model while minimising its complexity. An agent, according to the FEP, is then mandated to encode only relevant information to avoid unnecessary complex models that don't improve its performances.

The generative model in our agent represents a set of variables, $\mu_x = \{\mu_{l_1}, \mu_{l_2}, \mu_{m_1}, \mu_{m_2}\}$ acting as beliefs about the hidden causes $x$ of sensations $\rho = \{\rho_{l_1}, \rho_{l_2}, \rho_{m_1}, \rho_{m_2}\}$. In our implementation, these variables are not strongly related to how sensory data $\rho$'s are actually generated i.e. they contain no information about, for example, the details of the agent's body or environment. It could be argued that this agent does not even possess a generative model in its purest sense since it cannot generate predictions in line with sensory data. Our interpretation aligns with \cite{clark2015radical, bruineberg2016anticipating} in saying that, as already mentioned above, a generative model in the context of embodied agents should be assessed based on its ability to allow the agent to perform a task (action-oriented), rather than on how well it can reconstruct and accurately predict sensory input (perception-oriented).

In our investigations we first provided an account of phototaxis functionally consistent with vehicle 2b \citep{braitenberg1986vehicles}, and how it could be implemented under the FEP. To allow for such behaviour we endowed our agent with beliefs on exteroceptive readings (i.e. light intensity) and mapped them to proprioceptive ones (i.e. motor velocity) using simple linear contralateral relations. In addition, this agent needs to implement high and low precisions on exteroceptive and proprioceptive inputs respectively,  with precisions on their interaction placed somewhere in between. This agent performs phototaxis through active inference without explicitly encoding information about a target end-point in the generative model, i.e. it does not specify a light intensity to achieve. The beliefs' structure of this agent encodes a target ``state of affairs" rather than a final goal. This agent thus minimises its free energy by complying with this state of affairs, not by achieving an explicit goal state. Phototaxis is just a consequence of how priors relating light levels to motor velocities are implemented.

We then explored behaviours defined as ``pathological", inspired by work on the FEP for motor disorders, psychosis and schizophrenia among others \citep{adams2013psychosis, friston2016dysconnection}. Some pathologies, it has been suggested, can be recapitulated under the FEP when different weighting parameters in a generative model (precisions) are altered. Our simulations investigated a case where a combination of high precisions on proprioceptive prediction errors and low precisions on the relation between extero- and proprioceptive beliefs resulted in a reduced (or complete lack of) ability to move. A similar idea was presented in \cite{brown2013active} where decreasing the confidence (precision) of sensations about self-generated movements is thought to be a necessary condition for the initiation of action.

We finally explored how simple changes in priors can produce new emergent behaviours, for example an alternative version of phototaxis and two types of photophobia, in line with vehicles 2a and 3a-b \citep{braitenberg1986vehicles}.

The implementation of phototaxis we presented here is admittedly rather complicated. However, it constitutes a complete example of an agentive system operating on the basis of the FEP and we would argue that models like this will be vital if the potential of the FEP and active inference is to be fully understood. The action-oriented model we implemented has also begun to address some of the concerns about the FEP (\cite{clark2015radical, bruineberg2016anticipating}), essentially worrying that interpretations of this framework have been so far too dependent on a perception-oriented account of behaviour. We would argue that the FEP is neutral on the implementation details of agentive behaviour and is thus compatible with many different cognitive frameworks. Indeed this neutrality is perhaps a central strength of the FEP as it may provide a unified formalism within which to examine different understandings of cognitive systems.

\section{Future directions}
This presentation constitutes only a simple proof of concept of action-oriented approaches to the FEP, with a more complete analysis of our minimal generative models left for future work. In the future we will start by investigating the possible functional benefits of this architecture based on the FEP. For instance, preliminary results comparing standard Braitenberg vehicles and our implementation already suggest a higher robustness of the latter in very noisy environments because of low-pass filters implemented by \eqnref{eq:perception} and \eqnref{eq:action}. A second direction will address generative models with a less minimal set of assumptions, including for example prior beliefs for exteroceptive input (a target light intensity to achieve?) and the exploration of behaviours that can only be performed with the implementation of such mechanisms (maintaining a certain distance from the light source?). This will allow us to investigate the implications for the perception- vs. action-oriented debate on the interpretation of the FEP \citep{clark2015radical, bruineberg2016anticipating, allen2016cognitivism} for  more complex adaptive behaviours.

\footnotesize
\bibliographystyle{apalike}

\end{document}